# A Family of Interleaved High Step-Up DC-DC Converters by Integrating a Voltage Multiplier and an Active Clamp Circuits

R. Beiranvand, *Member, IEEE* and S. H. Sangani

*Abstract*—A family of interleaved current-fed high step-up dc-dc converters are introduced and analyzed here by combining a voltage multiplier (VM) and an active clamp circuit for high-voltage high-power applications. Low input currents and output voltages ripples values and high voltage-gains characteristics of these converters make them suitable for lots of dc-dc applications. All power devices operate entirely under soft switching conditions, even when wide load and input voltage variations are applied. Thus, they can be designed at high switching frequencies to reduce passive components sizes to achieve high-power density, one of the main targets of the power electronics researches. Also, their input and output ports common ground simplifies the gate-drives and control circuits. To verify the given analyses and simulations, a 120-320 V to 1 kV, 50-1300 W three-stage two-leg prototype converter has been implemented at 100 kHz. Based on the experimental results, maximum efficiency of 96.5% is achieved.

## I. INTRODUCTION

TRANSFORMER less high step-up converters are one of the most interesting topics in power electronics, due to their high power density. Recently, lots of researches have been done to introduce new switched-capacitor converters (SCCs) and to improve their performances and also to overcome some problems such as hard switching, voltage regulation, power rating, and voltage gain [1]- [6]. Also, pulsating and non-pulsating input and output currents of some ultra-high step-up converters has been analyzed in [7]. Integrating some quadratic boost converters and voltage-multiplier (VM) modules have been given in literature to achieve high-voltage gains which are suitable for low power applications [8]- [10]. Also, combining an interleaved boost stage with VMs has been suggested in [11]. Nowadays, there are lots of well-known VMs such as Cockcroft-Walton, Dickson, or their hybrid configurations [12]. But, some useful VM cells, introduced in [13] and [14], are not suitable for high-voltage high-power applications in their current forms. Also, some integrated VMs and coupled-inductor based dc-dc converters, with low voltage stresses on the components, have been introduced for high-voltage gain applications [15] and [16]. The VMs have some other applications such as equalizing series-connected energy storage cells voltages like batteries and super capacitors, too [17]. For achieving a high-voltage gain converter, another approach is finding in the literature without using coupled-inductor, transformer, VM, or multiple lifting techniques [18]. In summary, these step-up converters are not suitable for high-voltage and high-power applications. Consequently, some modifications and improvements are necessary.

Beside the desired high-voltage gain, soft switching capability is also important to reduce switching losses and EMI noises to allow high switching frequency operation to reduce passive components sizes to achieve high power density. Although, some existing topologies (for example [19] and [20]) cannot realize this feature to reduce the switching losses, but some researchers achieve zero-current-switching (ZCS) operation by integrating the switched-capacitor and coupled-inductor techniques [21], [22]. But, the ZCS technique doesn't remove dominant switching losses and EMI noises at MOSFETs turn-on moments in MOSFET-based converters, as compared to zero voltage switching (ZVS) approach [23].

Also, the proposed passive lossless clamped circuit in [24] is more flexible and versatile as compared to the classical passive lossless clamped circuits. Some active clamp circuits are employed in [25]– [28] to facilitate soft-switching performance for power switches in wide ranges of output power variations. But, some of these converters, for instance [28], are not suitable for high-voltage applications in their existing configurations.

On the other hand, although the given converter in [29] is suitable for high-power applications, but it cannot be used for high-voltage applications. This converter has been used at low switching frequencies, up to 40 kHz, by the authors to reduce its hard switching problems. This issue can be highlighted as one of the main disadvantages of the given converters in [30] and [31], too. Although, the given converters in [32] and [33] that use low voltage rating capacitors are suitable for high-voltage applications, but [33] has power density problem, due to its hard switching operation and its massive transformers. The combined SCC and switched-inductor in [34] is not suitable for high-power and high-voltage applications and there is no common ground between its ports, too. Although, the given converter by S. Sathyan et al., [35] has some good features such as soft switching operation and high-voltage high-power capabilities, but transformers with air gaps reduce its power density. Also, [36] is not suitable for high-voltage applications in its existing form.

As mentioned, nowadays some researchers are mainly focused on integrating SCCs or diode-based VM circuits, interleaved boost converters, and coupled-inductor techniques to improve voltage gain and to reduce input and output current ripples, and also to achieve low voltage stresses on the components. But, lots of the given topologies are not suitable for high-power and high-voltage applications at high switching frequencies. Here, a general topology of a family of high step-up dc-dc converter is proposed based on the Cockcroft-Walton VM circuit for these applications. Their all power devices soft switching operation, even under wide input voltage and load variations, make them suitable for high-frequency applications. Also, the input and output ports common ground simplifies the gate-driver and control circuits. Also, their interleaving capabilities reduce the input filter volume and output capacitors sizes, significantly.

General configuration of the proposed converters is given in Sec. II. Key waveforms and different operational states are given in Sec. III. The inductors currents are calculated in Sec. IV in general forms, and different time intervals durations and power MOSFETs duty cycles are derived in Sec. V. Also, voltage stresses on the capacitor of the active clamp circuit and the power MOSFETs are given in Sec. VI. Then, the diodes average currents values and the output dc current value are calculated in Sec. VII. Also, voltage ripples of the capacitors are derived in Sec. VIII, and ZVS operation under the worst-case conditions is discussed in Sec. IX. Experimental results, and conclusion are respectively described in sections X and XI. Finally, some different operational states are analyzed in Appendix, in brief.

## II. GENERAL CONFIGURATION OF THE PROPOSED CONVERTERS

A 3-stage 3-leg configuration of the proposed interleaved converters is given in Fig. 1. Depends on the application, this topology can easily be extended properly to m-stage n-leg configuration for high output voltage and high output power applications. Higher output voltages are obtained by increasing the stage-numbers, properly. Also, using more legs reduces the input current and output voltage ripples values by considering some proper phase shifts in between the legs gate-drive signals. A 3-stage 3-leg configuration of the proposed interleaved converters is given in Fig. 1. Depends on the application, this topology can easily be extended properly to m-stage n-leg configuration for high output voltage and high output power applications. Higher output voltages are obtained by increasing the stage-numbers, properly. Also, using more legs reduces the input current and output voltage ripples values by considering some proper phase shifts in between the legs gate-drive signals.

Finally, it must be mentioned that, the given converter is a general configuration that can be used for lots of high-voltage high-power applications. Also, it can be used for renewable applications that lots of renewable sources, such as PVs, are using, simultaneously. For the PV-based applications, each leg of the input port is connected to a single PV, separately. Lots of these legs can be connected easily in parallel, regardless of their components and input sources differences, because each leg is a pulsating current source at the output point of view. So, each leg duty ratio can be controlled separately to cover these differences, if exist. Therefore, the load current can be shared properly between the input sources by using proper control loops. These topics, in addition of the control approach, are not addressed here to shorten the subject. Also, all of the input ports are connected to a single input source here, and all duty ratios are the same, by considering proper phase-shifts in between, to simplify the analysis of the power section of the proposed converter.

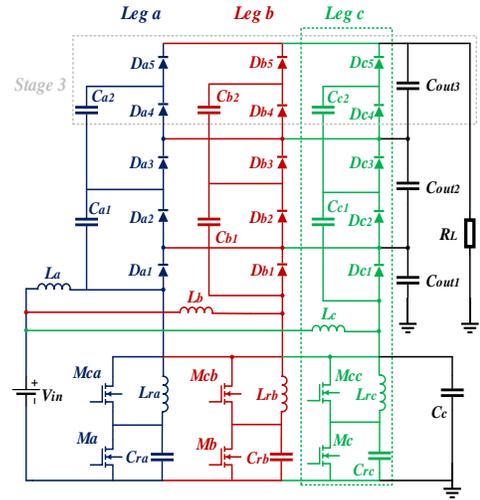

Fig. 1. A 3-stage 3-leg configuration of the proposed converter.

## III. Converter Key Waveforms and Operational States

The proposed converter in a three-stage two-leg configuration has been shown in Fig. 2, and its one leg key waveforms have been plotted in Fig. 3. The main aim of this paper is describing and analyzing the converter power section operation principles, without wrestling any other complex issues. Therefore, to simplify the discussions and analyses, both legs of the converter gate-signals are considered to be the same with only a constant 180-degree phase shifted in between to reduce the input current and output voltage ripples values. It must be mentioned that it also is possible to connect bottom pin of $C_{out1}$ capacitor in Fig. 2 to the common input source to reduce its voltage stress. But, connecting this pin to the power ground, as used here, reduces the input current ripple value, significantly. Consequently, it reduces the necessary input filter, but an extra more voltage stress, equal to $V_{in}$, is applied on $C_{out1}$. Of course, the other output capacitors voltage stresses are not different in both cases. The analyses show that the components have low current and voltage ratings. So, it has a small volume, in practice.

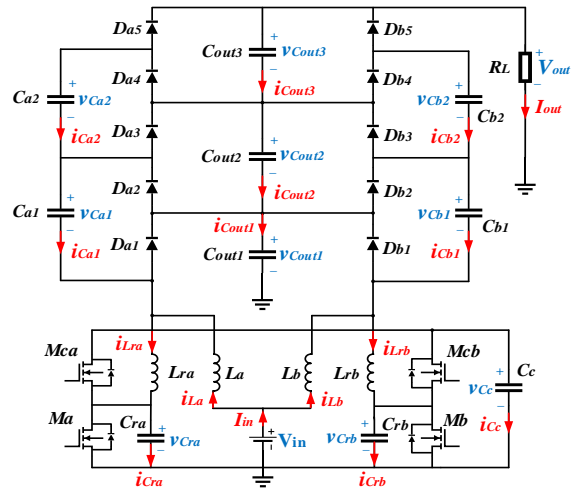

Fig. 2. A two-leg three-stage configuration of the converter.

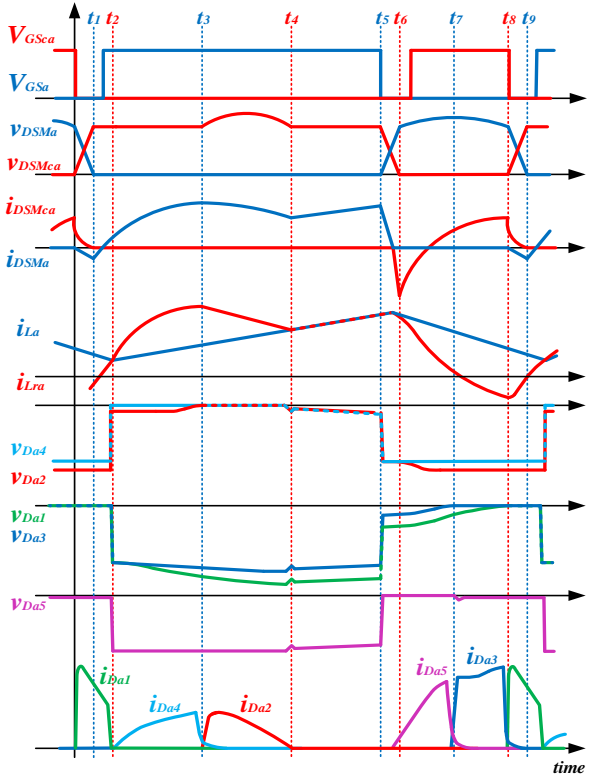

Fig. 3. Some key waveforms of a single leg of the given converter.

## IV. CALCULATING THE INDUCTORS CURRENTS

Inductors currents are key waveforms that describe the given converter behaviour and they can be calculated in canonical forms during each operational state. To simplify the analyses, we assume $C_{ra} \ll C_c \ll C_{out1} = C_{out2} = C_{out3}$ and $C_{a1} = C_{a2}$. In the same manners, as done in Appendix for some operational states, the other operational states equations can be derived, straightforwardly. A simple investigation shows that these equations can be expressed as follows:

$$\begin{bmatrix} L_a s & L_{ra}s + \frac{1}{C_{12\,i}}\frac{1}{s} \\ \left(\frac{1}{C_{12\,i}} - \frac{1}{C_{22\,i}}\right)\frac{1}{s} & L_{ra}s + \frac{1}{C_{22\,i}}\frac{1}{s} \end{bmatrix} \begin{bmatrix} I_{L_{a\,i}}(s) \\ I_{L_{ra\,i}}(s) \end{bmatrix} = \begin{bmatrix} \frac{a_i}{s} + b_i \\ \frac{c_i}{s} + d_i \end{bmatrix} \quad (1)$$

Here, $i \in \{1, 2, 3, \ldots, 8\}$ shows the $i^{th}$ operational state, and the exponential term, due to each operational state delay time, i.e., $t_i$ is ignored to simplify the analyses, as discussed in Appendix. Also, $C_{12\,i}$, is the equivalent capacitance seen from $L_{ra}$ inductor negative pin to the ground, and $C_{22\,i}$ is equivalent capacitance seen by $L_{ra}$ during the state $i$, and

$$\begin{cases} a_i = -v_{C_{ra}}(t_i) \\ b_i = V_{in}(s) + L_a i_{L_a}(t_i) + L_{ra} i_{L_{ra}}(t_i) \\ c_i = [\Gamma]_i [V_{L_a}(t_i)]^T \\ d_i = L_{ra} i_{L_{ra}}(t_i) \end{cases} \quad (2)$$

Here, $[\Gamma]_i$ is the $i^{th}$ row of the matrix $[\Gamma]$, defined as follows:

$$[\Gamma] = \begin{bmatrix} s & 0 & 0 & 0 & 0 & 0 & 0 \\ s & s & 0 & 0 & 0 & -1 & -1 \\ s & 0 & 0 & 0 & 0 & -1 & 0 \\ s & s & s & 0 & -1 & 0 & 0 \\ s & s & s & -1 & -1 & 0 & 0 \\ s & s & s & -1 & 0 & -1 & -1 \\ s & s & 0 & -1 & 0 & -1 & 0 \\ s & 0 & 0 & -1 & 0 & 0 & 0 \end{bmatrix} \quad (3)$$

Also, by using $V_{C_{out\,i}}$ term instead of $V_{C_{out\,i}}(s)$ for simplicity:

$$[V_{L_a}(t_i)] = \begin{bmatrix} V_{C_{out1}} & V_{C_{out2}} & V_{C_{out3}} & v_{C_{ra}}(t_i) & v_{C_p}(t_i) & v_{C_{a1}}(t_i) & v_{C_{a2}}(t_i) \end{bmatrix} \quad (4)$$

For configurations with more stages, numbers of rows and columns of matrix $[\Gamma]$ and number of columns of vector $[V_{L_a}(t_i)]$ must be changed, accordingly. Here,

$$1/C_{12\,i} = \begin{cases} 0 & ,\text{if } M_a/D_{M_a} \text{ are conducting} \\ \lambda_c/C_c & ,\text{if } M_{ca}/D_{M_{ca}} \text{ are conducting} \\ 1/C_{ra} & ,\text{otherwise} \end{cases} \quad (5)$$

$$\lambda_c = \frac{1}{1 + C_{ra}/C_c} \quad (6)$$

(5) can be rewritten as follows in matrix form:

$$[1/C_{12\,i}] = \begin{bmatrix} 0 & 0 & 0 & 0 & 1 & 0 & 0 & 1 \\ 0 & 0 & 0 & 0 & 0 & \lambda_c & \lambda_c & 0 \end{bmatrix}^T \begin{bmatrix} 1/C_c \\ 1/C_{ra} \end{bmatrix} \quad (7)$$

Where, $A^T$ is transposd of matrix A. Also, we can write:

$$[1/C_{22\,i}] = \begin{bmatrix} 0 & 0 & 0 & 0 & 0 & \lambda_c & \lambda_c & 0 \\ 0 & 0 & 0 & 0 & 1 & 0 & 0 & 1 \\ 0 & 0 & 0 & 1 & 1 & 0 & 0 & 0 \\ 0 & 1 & 1 & 0 & 0 & 1 & 1 & 0 \\ 0 & 1 & 0 & 0 & 0 & 1 & 0 & 0 \end{bmatrix}^T \begin{bmatrix} 1/C_c \\ 1/C_{ra} \\ 1/C_p \\ 1/C_{a1} \\ 1/C_{a2} \end{bmatrix} \quad (8)$$

Now, from (1) during each operational state we can write:

$$I_{L_{ai}}(s) = i_{L_a}(t_i) \frac{s^3 + \alpha_{1i}\omega_{1i}s^2 + \beta_{1i}\omega_{1i}^2 s + \gamma_{1i}\omega_{1i}^3}{(s^2 + \omega_{1i}^2)(s^2 + \omega_{2i}^2)} \quad (9)$$

$$\begin{cases} \omega_{1i} = \Omega_i \sqrt{1 + \sqrt{1 - \zeta_i^4/\Omega_i^4}}, \\ \omega_{2i} = \Omega_i \sqrt{1 - \sqrt{1 - \zeta_i^4/\Omega_i^4}}, \\ \Omega_i^2 = \frac{1}{2}\left(\frac{1}{L_{ra}C_{22\,i}} + \frac{1}{L_a C_{22\,i}} - \frac{1}{L_a C_{12\,i}}\right), \\ \zeta_i^4 = \frac{1}{L_{ra}L_a C_{12\,i}}\left(\frac{1}{C_{22\,i}} - \frac{1}{C_{12\,i}}\right). \end{cases} \quad (10)$$

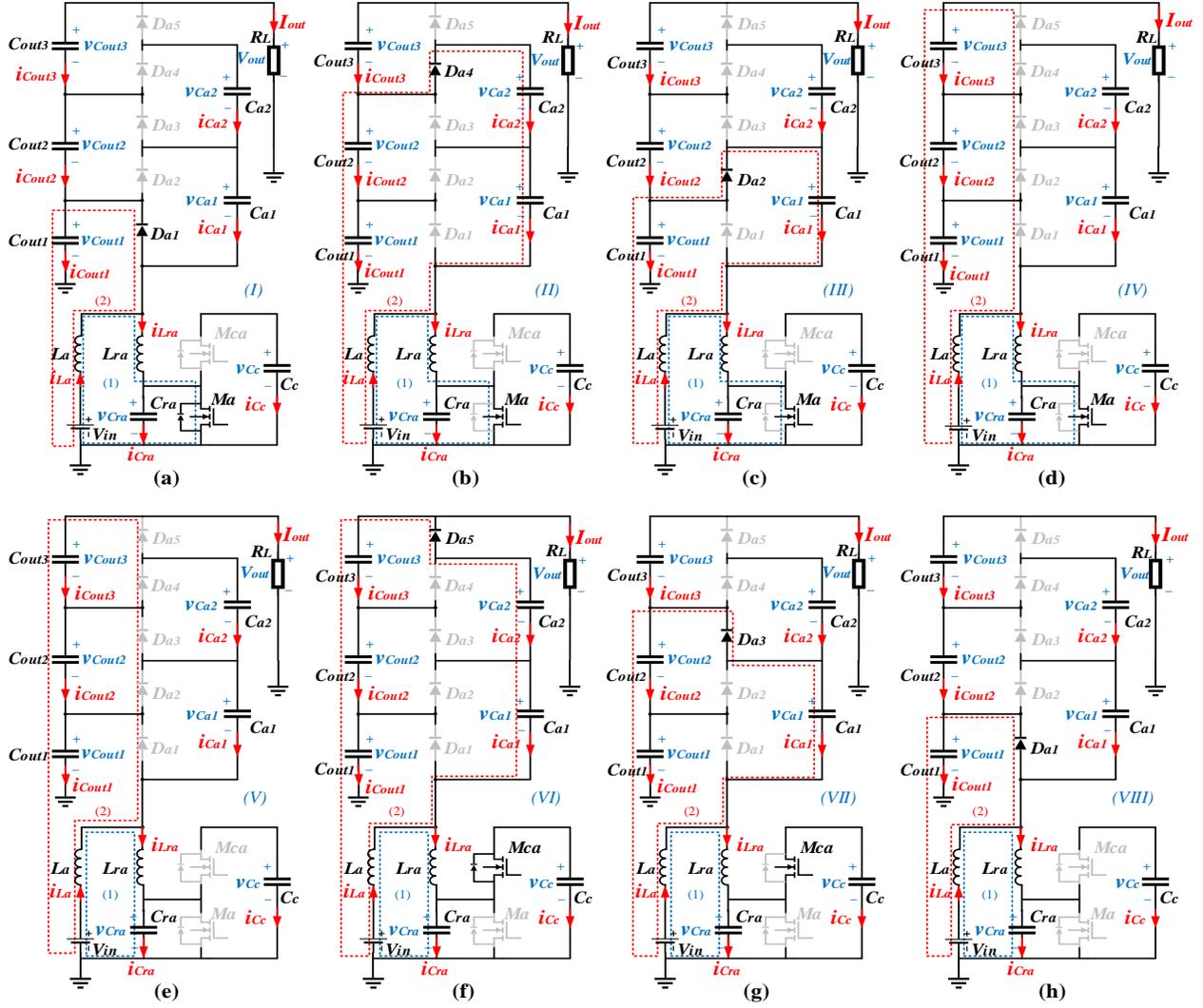

Fig. 4. Equivalent circuits of a single leg of the converter in different operational states during a switching period: (a) $D_{Ma}/M_a$, $D_{a1}$, (b) $M_a$, $D_{a4}$, (c) $M_a$, $D_{a2}$, (d) $M_a$, (e) all power devices are off, (f) $D_{Mca}/M_{ca}$, $D_{a5}$, (g) $M_{ca}$, $D_{a3}$, and (h) $D_{a1}$.

Also, other parameters are given as follows:

$$\begin{cases} \alpha_{1i} = \dfrac{a_i - c_i}{L_a \omega_{1i} i_{L_a}(t_i)} \\ \beta_{1i} = \dfrac{1}{L_a L_{ra} \omega_{1i}^2 i_{L_a}(t_i)} \left( \dfrac{b_i}{C_{22\,i}} - \dfrac{d_i}{C_{12\,i}} \right) \\ \gamma_{1i} = \dfrac{\beta_{1i}(a_i C_{12\,i} - c_i C_{22\,i})}{\omega_{1i}(b_i C_{12\,i} - d_i C_{22\,i})} \end{cases} \quad (11)$$

(9) can be expressed in standard form as follows:

$$I_{L_{ai}}(s) = i_{L_a}(t_i) \left( k_{1i} \dfrac{s + \omega_{1i}\cot\theta_{1i}}{s^2 + \omega_{1i}^2} + k_{2i} \dfrac{s + \omega_{2i}\cot\theta_{1i}}{s^2 + \omega_{2i}^2} \right) \quad (12)$$

This equation can be expressed in another proper form:

$$I_{L_{ai}}(s) = i_{L_a}(t_i) \left( \dfrac{k_{1i}}{\sin\theta_{1i}} \dfrac{\sin\theta_{1i}\, s + \omega_{1i}\cos\theta_{1i}}{s^2 + \omega_{1i}^2} + \dfrac{k_{2i}}{\sin\theta_{2i}} \dfrac{\sin\theta_{2i}\, s + \omega_{2i}\cos\theta_{2i}}{s^2 + \omega_{2i}^2} \right) \quad (13)$$

Now, $i_{L_a}$ during each operational state is generally given:

$$i_{L_{ai}}(t) = i_{L_a}(t_i) \left( \dfrac{k_{1i}}{\sin\theta_{1i}} \sin(\omega_{1i}(t - t_i) + \theta_{1i}) + \dfrac{k_{2i}}{\sin\theta_{2i}} \sin(\omega_{2i}(t - t_i) + \theta_{2i}) \right) \quad (14)$$

$$\begin{cases} k_{1i} = \dfrac{1 - \beta_{1i}}{1 - \lambda_{\omega i}^2}, \quad k_{2i} = \dfrac{\beta_{1i} - \lambda_{\omega i}^2}{1 - \lambda_{\omega i}^2}, \\ \theta_{1i} = \cot^{-1}\left[ \dfrac{\alpha_{1i} - \gamma_{1i}}{k_{1i}(1 - \lambda_{\omega i}^2)} \right], \\ \theta_{2i} = \cot^{-1}\left[ \dfrac{\gamma_{1i} - \alpha_{1i}\lambda_{\omega i}^2}{k_{2i}\lambda_{\omega i}(1 - \lambda_{\omega i}^2)} \right]. \end{cases} \quad (15)$$

Where, $\lambda_{\omega i} = \omega_{2i}/\omega_{1i}$. From (1), $I_{L_{ra}}(s)$ can be expressed as follows during each operational state, in the same manner:

$$I_{L_{ra\,i}}(s) = i_{L_{ra}}(t_i) \dfrac{s^3 + \alpha_{2i}\omega_{1i}s^2 + \beta_{2i}\omega_{1i}^2 s + \gamma_{2i}\omega_{1i}^3}{(s^2 + \omega_{1i}^2)(s^2 + \omega_{2i}^2)} \quad (16)$$

$$\begin{cases} \alpha_{2i} = \dfrac{c_i}{L_{ra} i_{L_{ra}}(t_i)\omega_{1i}}, \quad \gamma_{2i} = \dfrac{a_i \beta_{2i}}{b_i \omega_{1i}}, \\ \beta_{2i} = \left( \dfrac{1}{C_{22\,i}} - \dfrac{1}{C_{12\,i}} \right) \dfrac{b_i}{L_{ra} L_a i_{L_{ra}}(t_i)\omega_{1i}^2}. \end{cases} \quad (17)$$

As done recently, (16) can be rewritten as follows, too:

$$I_{L_{ra\,i}}(s) = i_{L_{ra}}(t_i) \left( \frac{k_{3i}}{\sin\theta_{3i}} \frac{\sin\theta_{3i}\, s + \omega_{1i}\cos\theta_{3i}}{s^2 + \omega_{1i}^2} \right.$$
$$\left. + \frac{k_{4i}}{\sin\theta_{4i}} \frac{\sin\theta_{4i}\, s + \omega_{2i}\cos\theta_{4i}}{s^2 + \omega_{2i}^2} \right) \quad (18)$$

$$\begin{cases} k_{3i} = \dfrac{1-\beta_{2i}}{1-\lambda_{\omega i}^2}, \quad k_{4i} = \dfrac{\beta_{2i}-\lambda_{\omega i}^2}{1-\lambda_{\omega i}^2} \\ \theta_{3i} = \cot^{-1}\left[\dfrac{\alpha_{2i}-\gamma_{2i}}{k_{3i}(1-\lambda_{\omega i}^2)}\right], \\ \theta_{4i} = \cot^{-1}\left[\dfrac{\gamma_{2i}-\alpha_{2i}\lambda_{\omega i}^2}{k_{4i}\lambda_{\omega i}(1-\lambda_{\omega i}^2)}\right]. \end{cases} \quad (19)$$

Therefore, we can write:

$$i_{L_{ra\,i}}(t) = i_{L_{ra}}(t_i) \left( \frac{k_{3i}}{\sin\theta_{3i}} \sin(\omega_{1i}(t-t_i)+\theta_{3i}) \right.$$
$$\left. + \frac{k_{4i}}{\sin\theta_{4i}} \sin(\omega_{2i}(t-t_i)+\theta_{4i}) \right) \quad (20)$$

During special states I and IV, (13) and (20) can be more simplified as follows, because some coefficients are zero:

$$\begin{cases} i_{L_{a1}}(t) = \dfrac{V_{in} - V_{C_{out1}}}{L_a}(t-t_1) + i_{L_a}(t_1) \\ i_{L_{ra1}}(t) = \dfrac{V_{C_{out1}}}{L_{ra}}(t-t_1) + i_{L_{ra}}(t_1) \end{cases} \quad (21)$$

$$i_{L_{ra4}}(t) = i_{L_{a4}}(t) = \frac{V_{in}}{L_a + L_{ra}}(t-t_4) + i_{L_a}(t_4) \quad (22)$$

## V. Calculating the Different Time Intervals and the Power MOSFETs Duty Cycles

Now, we are ready to do more algebraic calculations to identify each subinterval duration, as well as the power MOSFETs duty cycles. Considering Fig. 3, state I is finished at $t_2$, when $i_{L_{ra\,1}}(t) = i_{L_{a\,1}}(t)$. Thus, the first subinterval duration can be identified easily, by considering (21).

$$t_s = t_2 - t_1 = \frac{i_{L_a}(t_1) - i_{L_{ra}}(t_1)}{V_{C_{out1}}\left(\dfrac{1}{L_a}+\dfrac{1}{L_{ra}}\right) - \dfrac{V_{in}}{L_a}}$$
$$= \frac{1}{\omega_r \lambda_L} \frac{j_{L_a\,mx} - j_{L_{ra}\,min}}{\dfrac{M}{m}\left(1+\dfrac{1}{\lambda_L}\right) - 1} \quad (23)$$

Summation of the time-durations of the operational states I-IV can be obtained by considering Fig. 3 and having $M_a$ MOSFET duty cycle.:

$$t_5 - t_1 = D_{M_a} T_s \quad (24)$$

More calculations are necessary to identify the time duration of the subinterval V, i.e., $t_6 - t_5$. At the beginning of this subinterval, $v_{C_{ra}}(t_5) = 0$ and $i_{L_a}(t_5) = i_{L_{ra}}(t_5)$. Considering (10), (11), and (13) or from (14), (20) and Fig. 4 (e), inductors currents are easily identifying during this operational state:

$$i_{L_{ra5}}(t) = i_{L_{a5}}(t) = \frac{i_{L_{ra}}(t_5)}{\sin\theta_{35}} \sin(\omega_{15}(t-t_5)+\theta_{35}) \quad (25)$$

$$\begin{cases} \theta_{35} = \cot^{-1}\left(\dfrac{V_{in}}{i_{L_{ra}}(t_5)Z_r}\sqrt{\dfrac{\lambda_L}{1+\lambda_L}}\right), \\ \omega_{15} = \omega_r \sqrt{\dfrac{\lambda_L}{1+\lambda_L}}, \quad \lambda_L = \dfrac{L_{ra}}{L_a}. \end{cases} \quad (26)$$

Now, the voltage on $C_{ra}$ during this operational state is obtained by considering Fig. 4 (e), (25), and doing some straightforward algebraic calculations:

$$v_{C_{ra5}}(t) = v_{C_{ra}}(t_5) + \frac{1}{C_{ra}}\int_{t_5}^{t} i_{C_{ra}}(t)dt =$$
$$\frac{i_{L_{ra}}(t_5)}{\omega_{15}C_{ra}\sin\theta_{35}}\left(\cos\theta_{35} - \cos(\omega_{15}(t-t_5)+\theta_{35})\right) \quad (27)$$

From Fig. 3, this operational state is terminated when $v_{C_{ra}}(t_6) = V_{C_c}$. Thus, by considering (27) we have:

$$t_r = t_6 - t_5 = \frac{1}{\omega_{15}}(\cos^{-1}\gamma_5 - \theta_{35}) \quad (28)$$

$$\gamma_5 = \cos\theta_{35} - \sqrt{\frac{\lambda_L}{1+\lambda_L}} \frac{V_{C_c}\sin\theta_{35}}{i_{L_{ra}}(t_5)Z_r} \quad (29)$$

Summation of the time-durations of the operational states VI and VII can be obtained easily by considering Fig. 3 and having $M_{ca}$ MOSFET duty cycle, i.e., $D_{M_{ca}}$ as follows:

$$t_8 - t_6 = D_{M_{ca}} T_s \quad (30)$$

At the beginning of the subinterval VIII, $v_{C_{ra}}(t_8) = V_{C_c}$. Considering (14), (20) and Fig. 4 (h), as discussed earlier for the subinterval V, the inductors currents can be identifying during the operational state VIII as follows:

$$\begin{cases} i_{L_{a8}}(t) = i_{L_a}(t_8) + \dfrac{V_{in}-V_{C_{out1}}}{L_a}(t-t_8) \\ i_{L_{ra8}}(t) = \dfrac{i_{L_{ra}}(t_8)}{\sin\theta_{38}}\sin(\omega_{18}(t-t_8)+\theta_{38}) \end{cases} \quad (31)$$

$$\theta_{38} = \cot^{-1}\left(\frac{V_{C_{out1}} - v_{C_{ra}}(t_8)}{i_{L_{ra}}(t_8)Z_r}\right), \quad \omega_{18} = \omega_r \quad (32)$$

So, the applied voltage on $C_{ra}$ is obtained by considering (31), Fig. 4 (h), and doing some algebraic calculations:

$$v_{C_{ra8}}(t) = v_{C_{ra}}(t_8) + \frac{1}{C_{ra}}\int_{t_8}^{t} i_{L_{ra8}}(t)dt =$$
$$V_{C_c} + i_{L_{ra}}(t_8)\frac{\cos\theta_{38} - \cos(\omega_{18}(t-t_8)+\theta_{38})}{\omega_{18}C_{ra}\sin\theta_{38}} \quad (33)$$

From Fig. 3, this operational state is terminated when $C_{ra}$ is completely discharged, i.e., $v_{C_{ra}}(t_9) = 0$. Thus, this time interval is obtained by equating (33) to be zero which leads to:

$$t_f = t_9 - t_8 = \frac{1}{\omega_{18}}(\cos^{-1}\gamma_8 - \theta_{38}) \quad (34)$$

$$\gamma_8 = \cos\theta_{38} + \frac{V_{C_c}\sin\theta_{38}}{i_{L_{ra}}(t_8)Z_r} \quad (35)$$

Now, from Fig. 3 a complete switching period is obtained:

$$T_s = t_9 - t_1 = \\ (t_5 - t_1) + (t_6 - t_5) + (t_8 - t_6) + (t_9 - t_8) \quad (36)$$

Also, duty cycle of the auxiliary switch can be expressed as follows by substituting (24), (28), (30), and (34) into (36).

$$D_{M_{ca}} = 1 - D_{M_a} - \frac{F}{2\pi}\left[\sqrt{1+1/\lambda_L}\,(\cos^{-1}\gamma_5 - \theta_{35}) + \cos^{-1}\gamma_8 - \theta_{38}\right] \quad (37)$$

## VI. CALCULATING THE VOLTAGES STRESSES ON THE CAPACITOR OF THE ACTIVE CLAMP CIRCUIT AND THE POWER MOSFETs

Voltages stresses on the capacitor of the clamp circuit and the power MOSFETs are calculated, here. Considering the loop containing, $V_{in\,a}$, $L_a$, $L_{ra}$, and $C_{ra}$, in Fig. 2 or loop (1) in Fig. 4, and averaging the given KVL equation over a switching period, we can write:

$$\langle v_{L_a}(t)\rangle_{T_s} + \langle v_{L_{ra}}(t)\rangle_{T_s} + \langle v_{C_{ra}}(t)\rangle_{T_s} - \langle V_{in\,a}(t)\rangle_{T_s} = 0 \quad (38)$$

Also, by considering the volt-second balance principle during a switching period under the steady state condition for both inductors, (38) can be more simplified:

$$\langle v_{C_{ra}}(t)\rangle_{T_s} = \frac{1}{T_s}\int_0^{T_s} v_{C_{ra}}(t)dt = V_{in\,a} \quad (39)$$

(39) is easily calculated by considering $v_{C_{ra}}(t)$ during each subinterval separately, as calculated in the last section, for the operational states V and VIII. During $t_i \le t < t_{i+1}$ subinterval, for $i \in \{1,2,3,4\}$, voltage across $C_{ra}$, connected in paralleled to the drain-source of $M_a$, is essentially zero:

$$v_{C_{ra}i}(t) = 0 \qquad i \in \{1,2,3,4\} \quad (40)$$

But, for $i \in \{5,6,7,8\}$, this voltage can be calculated by considering (20) as follows:

$$v_{C_{ra}i}(t) = v_{C_{ra}}(t_i) + \frac{i_{L_{ra}}(t_i)}{C_{DSi}}\left(\frac{k_{3i}}{\omega_{1i}}\cot\theta_{3i} + \frac{k_{4i}}{\omega_{2i}}\cot\theta_{4i}\right) \\ -\frac{i_{L_{ra}}(t_i)}{C_{DSi}}\left[\frac{k_{3i}}{\omega_{1i}\sin\theta_{3i}}\cos(\omega_{1i}(t-t_i)+\theta_{3i}) \right. \\ \left. +\frac{k_{4i}}{\omega_{2i}\sin\theta_{4i}}\cos(\omega_{2i}(t-t_i)+\theta_{4i})\right] \quad (41)$$

$$C_{DSi} = \begin{cases} C_{ra} & i \in \{5,8\} \\ C_{ra}+C_c & i \in \{6,7\} \end{cases} \quad (42)$$

Substituting (40) and (41) into (39), we can write:

$$V_{in\,a} = \sum_{i=5}^{8}\left\{\left[v_{C_{ra}}(t_i) + \frac{i_{L_{ra}}(t_i)}{C_{DSi}}\left(\frac{k_{3i}}{\omega_{1i}}\cot\theta_{3i} + \frac{k_{4i}}{\omega_{2i}}\cot\theta_{4i}\right)\right] \right. \\ \times \frac{t_{i+1}-t_i}{T_s} + \frac{i_{L_{ra}}(t_i)}{T_s C_{DSi}}\left[\frac{k_{3i}}{\omega_{1i}^2}\left(\frac{\sin(\omega_{1i}(t_{i+1}-t_i)+\theta_{3i})}{\sin\theta_{3i}}-1\right) \right. \\ \left.\left. +\frac{k_{4i}}{\omega_{2i}^2}\left(\frac{\sin(\omega_{2i}(t_{i+1}-t_i)+\theta_{4i})}{\sin\theta_{4i}}-1\right)\right]\right\} \quad (43)$$

In practice, $C_{ra} \ll C_c$ has been chosen. So, variations of $v_{C_{ra}}(t)$, which are very small during the operational states VI and VII, can be ignored to simplify the analyses. Considering this simplifying assumption, substituting (27), (30), (33), and (40) into (39), and doing some straightforward algebraic calculations, $V_{C_c}$ is approximated as:

$$V_{C_c} = \frac{1}{D_{M_{ca}}+f_s t_f}\left\{V_{in\,a} - \frac{i_{L_{ra}}(t_5)}{\omega_{15}C_{ra}}\left[t_r f_s \cot\theta_{35} + \frac{1}{\omega_{15}}\times\right.\right. \\ \left.\left(\frac{\sin(\omega_{15}t_r+\theta_{35})}{\sin\theta_{35}}-1\right)\right] + \frac{i_{L_{ra}}(t_8)}{\omega_{18}C_{ra}}\times \\ \left.\left[t_f f_s \cot\theta_{38} + \frac{1}{\omega_{18}}\left(\frac{\sin(\omega_{18}t_r+\theta_{38})}{\sin\theta_{38}}-1\right)\right]\right\} \quad (44)$$

In a very special case, where the rise and fall times are very smaller than a switching period, (44) can be simplified as follows which itself can also be identified by inspection easily under these conditions:

$$V_{C_c} = \frac{V_{in\,a}}{D_{M_{ca}}} = \frac{V_{in\,a}}{1-D_{M_a}}, \quad \text{if } t_f \ll T_s \text{ and } t_r \ll T_s \quad (45)$$

During $t_6 \le t \le t_7$, $i_{L_{ra}}(t)$ charges $C_{ra}$ and $C_c$ capacitors. Therefore, these paralleled capacitors are fully charged when $i_{L_{ra}}(t)$ current passes through zero. This zero crossing moment is identified by using a simple iterative method which is given by equating (20) to be zero.

$$t_x = t_6 - \frac{\theta_{36}}{\omega_{16}} - \\ \frac{1}{\omega_{16}}\sin^{-1}\left[\frac{k_{46}}{k_{36}}\frac{\sin\theta_{36}}{\sin\theta_{46}}\sin(\omega_{26}(t_x-t_6)+\theta_{46})\right] \quad (46)$$

Thus, maximum value of the applied voltage on the clamp circuit capacitor is given by considering (20), (46), and doing some more calculations:

$$V_{C_c\,max} = V_{C_{ra}}(t_6) + \frac{1}{C_c+C_{ra}}\int_{t_6}^{t_x} i_{L_{ra}}(t)dt \\ = V_{C_c} + \frac{\tau_{36}+\tau_{46}}{C_c+C_{ra}}i_{L_{ra}}(t_6) \quad (47)$$

$$\begin{cases}\tau_{36} = \dfrac{k_{36}}{\omega_{16}}\left(\cot\theta_{36} - \dfrac{\cos(\omega_{16}(t_x-t_6)+\theta_{36})}{\sin\theta_{36}}\right) \\ \tau_{46} = \dfrac{k_{46}}{\omega_{26}}\left(\cot\theta_{46} - \dfrac{\cos(\omega_{26}(t_x-t_6)+\theta_{46})}{\sin\theta_{46}}\right)\end{cases} \quad (48)$$

Finally, it must be mentioned that this calculated voltage is also directly applied across both power MOSFETs in each leg.

## VII. CALCULATING THE DIODES AVERAGE CURRENTS AND THE OUTPUT DC CURRENT VALUES

Averaging the given KCL equation at the output node of the converter, see Fig 2 (a), over a switching period leads to:

$$\langle i_{D_{a5}}(t)\rangle_{T_s} + \langle i_{D_{b5}}(t)\rangle_{T_s} - \langle i_{C_{out3}}(t)\rangle_{T_s} \\ -\langle i_{out}(t)\rangle_{T_s} = 0 \quad (49)$$

It must be mentioned that the dc current value of each capacitor is equal to zero under the steady state condition, due to the charge-balance principle. Therefore, it can be proved generally in the same manner that all diodes of the converter have the same average currents values which are equal to the output dc current value under the steady state condition divided by the number of the paralleled legs of the converter.

$$\langle i_{D_{aj}}(t)\rangle_{T_s} = \langle i_{D_{bj}}(t)\rangle_{T_s} = \cdots = I_{out}/n \quad (50)$$

Here, $j \in \{1,2,3,4,5\}$. So, output dc current is given by considering (50) and Fig. 4 (f), as follows:

$$I_{out} = \langle i_{D_{a1}}(t)\rangle_{T_s} + \langle i_{D_{b1}}(t)\rangle_{T_s} = \frac{2t_s}{T_s}\left[i_{L_a}(t_1) - i_{L_{ra}}(t_1)\right.$$
$$\left. + \frac{t_s}{2L_a}\left(V_{in} - \left(1 + \frac{L_{ra}}{L_a}\right)V_{C_{out1}}\right)\right]$$
$$+ \frac{2t_f}{T_s}\left(i_{L_a}(t_8) + \frac{V_{in} - V_{C_{out1}}}{2L_a}t_f\right)$$
$$+ \frac{2}{T_s}\frac{i_{L_{ra}}(t_8)}{\omega_{18}}\left(\frac{\cos(\omega_{18}t_f + \theta_{38})}{\sin\theta_{38}} - \cot\theta_{38}\right) \quad (51)$$

Now, the converter voltage gain can be identified as follows:

$$M = \frac{V_{out}}{V_{in}} = \frac{R_L I_{out}}{V_{in}} =$$
$$\frac{\lambda_L(\alpha_f^2 - \alpha_s^2) + 2\alpha_f j_{L_a\,max} + 2M_{C_c}}{\frac{2\pi}{F_n R_n} + \frac{\lambda_L}{m}\alpha_f^2 - \left(\frac{2+\lambda_L - \lambda_L^2}{m}\right)\alpha_s^2} \leq \frac{m}{1 - D_{Ma}} \quad (52)$$

$$\begin{cases} \alpha_r = \omega_r t_r, & \alpha_s = \omega_r t_s, & R_n = R_L/Z_r \\ M_{C_c} = V_{C_c}/V_{in}, & F_n = f_s/f_r, & j_{L_a\,max} = i_{L_a\,max} Z_r/V_{in} \end{cases} \quad (53)$$

Fig. 5 shows the voltage gain and the normalized output capacitors voltages values versus the duty cycle for $R_L$=1 kΩ and two different number of stages of the converter.

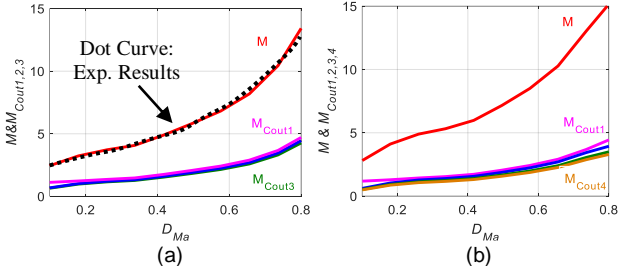

Fig. 5. Voltage gain and normalized output capacitors voltages versus duty cycle for different stages number $R_L$=1 kΩ (a) m=3 and (b) m=4.

## VIII. CALCULATING CAPACITORS VOLTAGES RIPPLES

Fig. 6 illustrates the converter different resonant capacitors voltages and currents waveforms during a switching period. Variations of the voltages of the converter capacitors are calculated, here. Considering Fig. 6, the applied voltage on $C_{a1}$ varies from its minimum to maximum value during the operational states II and III. Therefore, from Fig. 6 and by considering (50) and (51) this voltage variation is easily given:

$$v_{C_{a1}\,max} - v_{C_{a1}\,min} = \frac{1}{C_{a1}}\int_{t_2}^{t_4} i_{C_{a1}}(t)dt =$$
$$\frac{2}{C_{a1}}\int_{t_2}^{t_3}\left(i_{L_{ra}}(t) - i_{L_a}(t)\right)dt = \frac{V_{out}}{f_s C_{a1} R_L} \quad (54)$$

Also, $v_{C_{a2}}$ varies from its minimum to maximum value during the operational state II. Therefore, from Fig. 6 and by considering (50) this voltage variation is identified, too.

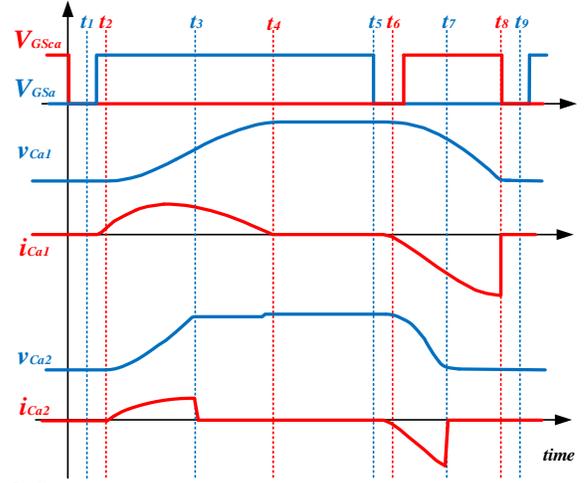

Fig. 6. Converter resonant stage capacitors simulated typical voltages and currents waveforms.

$$v_{C_{a2}\,max} - v_{C_{a2}\,min} = \frac{1}{C_{a2}}\int_{t_2}^{t_3} i_{C_{a2}}(t)dt =$$
$$\frac{1}{C_{a2}}\int_{t_2}^{t_3}\left(i_{L_{ra}}(t) - i_{L_a}(t)\right)dt = \frac{V_{out}}{2f_s C_{a2} R_L} \quad (55)$$

In the same manner and by considering Fig. 7, $v_{C_{out3}}$ varies from its minimum to maximum value during the operational state VI. Therefore, we can write:

$$v_{C_{out3}\,max} - v_{C_{out3}\,min} =$$
$$\frac{1}{C_{out3}}\int_{t_6}^{t_7}\left(i_{L_a}(t) - i_{L_{ra}}(t) - \frac{V_{out}}{R_L}\right)dt$$
$$= \left(\frac{1}{2f_s} + t_6 - t_7\right)\frac{V_{out}}{R_L C_{out3}} \quad (56)$$

Also, $v_{C_{out2}}$ varies from its minimum to maximum value during the operational states VI and VII. Therefore,

$$v_{C_{out2}\,max} - v_{C_{out2}\,min}$$
$$= \frac{1}{C_{out2}}\int_{t_6}^{t_8}\left(i_{L_a}(t) - i_{L_{ra}}(t) - \frac{V_{out}}{R_L}\right)dt$$
$$= \left(\frac{1}{f_s} + t_6 - t_8\right)\frac{V_{out}}{R_L C_{out2}} \quad (57)$$

Calculating voltage ripple of $C_{out1}$ is not straightforward, as done for other capacitors. It needs more algebraic calculations and Fig. 7 must be considered again. So, we have:

$$v_{C_{out1}\,max} - v_{C_{out1}\,min} = v_{C_{out1}}(t_y) - v_{C_{out1}}(t_z) =$$
$$-\frac{1}{C_{out1}}\int_{t_y}^{t_2} i_{C_{out1}}(t)dt - \frac{1}{C_{out1}}\int_{t_y}^{t_6 - T_s/2} i_{C_{out1}}(t)dt$$
$$-\frac{1}{C_{out1}}\int_{t_6 - T_s/2}^{t_z} i_{C_{out1}}(t)dt \quad (58)$$

Where, $i_{C_{out1}}$ during each time interval is given as follows:

$$\begin{cases} i_{L_{a1}}(t) - i_{L_{ra1}}(t) - I_{out}, & t_y \leq t < t_2 \\ i_{L_{a2}}(t) - i_{L_{ra2}}(t) - I_{out}, & t_2 \leq t < t_6 - T_s/2 \\ i_{L_{a3}}(t) - i_{L_{ra3}}(t) - I_{out} + & \\ \quad i_{L_{b3}}(t) - i_{L_{rb3}}(t), & t_6 - T_s/2 \leq t < t_z \end{cases} \quad (59)$$

Considering (14) and (20), and equating $i_{C_{out1}}$ to be zero during the operational states I and III, $t_y$ and $t_z$ are identified respectively by using the iterative method, in the same manners, as discussed in Sec. VIII for calculating $t_x$:

$$\begin{cases} i_{L_{a1}}(t_y) - i_{L_{ra1}}(t_y) - \dfrac{V_{out}}{R_L} = 0 \\ i_{L_{a3}}(t_z) - i_{L_{ra3}}(t_z) - \dfrac{V_{out}}{R_L} + i_{L_{b3}}(t_z) - i_{L_{rb3}}(t_z) = 0 \end{cases} \quad (60)$$

$$\begin{cases} i_{L_{b3}}(t_z) = i_{L_a}(t_z - T_s/2) = i_{L_{a6}}(t_z - t_6) \\ i_{L_{br3}}(t_z) = i_{L_{ra}}(t_z - T_s/2) = i_{L_{ra6}}(t_z - t_6) \end{cases} \quad (61)$$

Although $t_y$ can be approximated here by $t_2$, its exact value can be obtained from (60) using an iterative approach, easily.

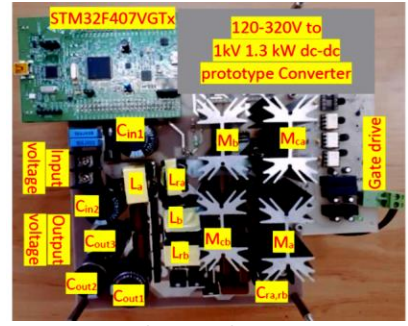

Fig. 8. Prototype converter photograph.

converter some key parameters and components specifications with some references. From table II, passive components are significantly reduced even when a common 100 kHz switching frequency is used. Therefore, using higher switching frequencies increases the power density, more effectively.

The resonant inductors inductances values are much smaller

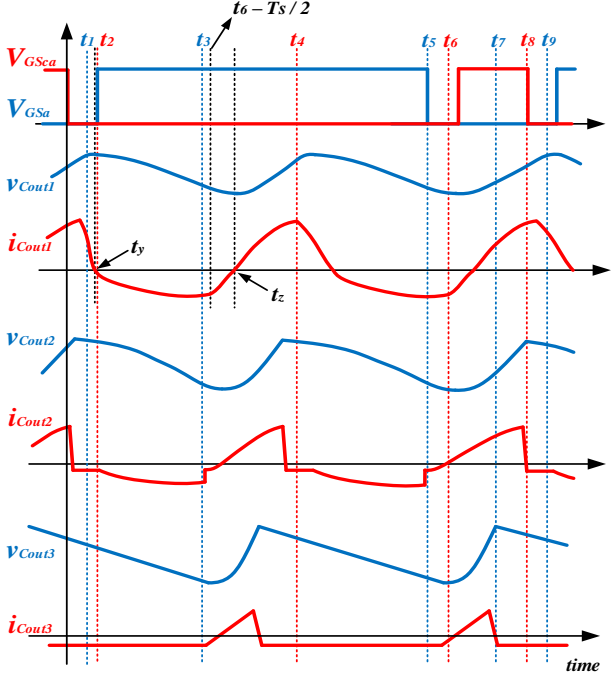

Fig. 7. Converter output stage capacitors magnified voltages ripples and currents waveforms.

### IX. ZVS OPERATION OF THE CONVERTER

To realize ZVS operation of the converter under the worst case-conditions, i.e., when maximum input voltage is applied to the converter under the light load conditions, (28) and (34) time intervals must be smaller than the given minimum dead time value. So, the following inequalities must be satisfied:

$$\begin{cases} \sqrt{1 + 1/\lambda_L} (\cos^{-1} \gamma_5 - \theta_{35}) \le \omega_r \Delta T_{min} \\ (\cos^{-1} \gamma_8 - \theta_{38}) \le \omega_r \Delta T_{min} \end{cases} \quad (62)$$

Here, $V_{in\,max}$ and $i_{L_{ra\,min}}(t_5)$ must be used to calculate $\theta_{35}$ and $\gamma_5$ by using (26) and (29), respectively. Also, $i_{L_{ra\,min}}(t_8)$ must be used to calculate $\theta_{38}$ and $\gamma_8$ by considering (32) and (35). Under the other conditions, $L_{ra}$ inductor current can fully charge or discharge $C_{ra}$ capacitor. Consequently, the MOSFETs body diodes are turned on and the ZVS conditions are well provided for turning on the MOSFETs.

### X. EXPERIMENTAL RESULTS

Fig. 8 gives the prototype converter photograph. Also, Tables I and II respectively show and compare a two-leg than the input inductances, as tabulated in Tables I and II, clearly. But, the input inductors currents ripples values are much smaller than the resonant inductors ones, as shown in Fig. 3. This problem causes more core losses per volume in the resonant inductors, in comparison to the input inductors ones. Consequently, the input to resonant inductors volumes ratio is not reduced proportionally by their inductances values, even if the same core materials be used to properly limit their maximum temperature, in practice. In addition of this problem, there are some free spaces here in the winding windows sections of both $L_{ra}$ and $L_{rb}$ in the implemented prototype converter, illustrated in Fig. 8. This fact shows that smaller inductors cores can be used. But, this optimization has been ignored, here. Because, it is not the main target of the current research. Of course, this optimization must be done properly for industrial applications, practically. Because, it affects the converter size and volume, power dissipations, efficiency and power density, as well as its cost.

TABLE I SPECIFICATIONS OF THE PROTOTYPE CONVERTER

| Parameters | Specifications |
| --- | --- |
| Micro controller | STM32F407VGTx |
| Gate driver | hcpl 3120 |
| Dead time | 350 ns |
| $M_{a,\,b}$ and $M_{ca,\,cb}$ | IRFP460Y49K |
| $D_{a1-5}$ and $D_{b1-5}$ | LT 6207 STPR1640CT |
| $L_r$ (µH) | 5 (EE2020, 4 turns, Litz: $0.3 \times 15$) |
| $L$ (µH) | 50 (PQ2020, 12 turns, Litz: $0.3 \times 15$) |
| $C_c$ (nF) | 100 nF1kV+1000nF 630V |
| $C_r$ (nF) | 3.3nF 630V |
| $C_{a1,2}$ and $C_{b1,2}$ (nF) | 330nF 630V |

TABLE II COMPARING SPECIFICATIONS OF THE MOST SIMILAR CONVERTERS

| Parameters | Ref. [39] | Ref. [40] | Ref. [41] | Proposed converter |
| --- | --- | --- | --- | --- |
| Input voltage (V) | 14 | 50 | - | 120-320 |
| Output voltage (V) | 42 | 300 | - | 1000 |
| Output power (W) | 10-140 | 20-140 | - | 50-1300 |
| $f_s$ (kHz) | 10-70 | 100 | 5/7 | 100 |
| L (µH) | 70 | 1330 | 25000 | 50 |
| $C_{a1}$ (µF) | 6600 | 100 µF+2 µF | 12000 | 0.33 |
| $C_{a2}$ (µF) | - | 100 µF+2 µF | 12000 | 0.33 |
| $C_{out1}$ (µF) | 4700 | 100 µF+2 µF | 12000 | 100 |
| $C_{out2}$ (µF) | 4700 | 100 µF+2 µF | 12000 | 100 |
| $C_{out3}$ (µF) | - | 100 µF+2 µF | 12000 | 100 |

Different experimental results are given in Figs. 9-13 under the two different worst-case conditions, i.e., (a) $V_{in\ min}$=120 V and $R_{L\ min}$=750 Ω, and (b) $V_{in\ max}$=320 V and $R_{L\ max}$=20 kΩ, to achieve $V_{out}$=1 kV. Figs. 9 and 10 show power MOSFETs duty cycles variations to regulate the output voltage when wide input voltage and load variations are applied.

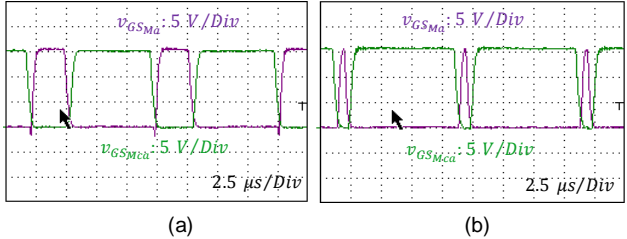
(a) (b)
Fig. 9. Gate drive signals under the two worst-case conditions.

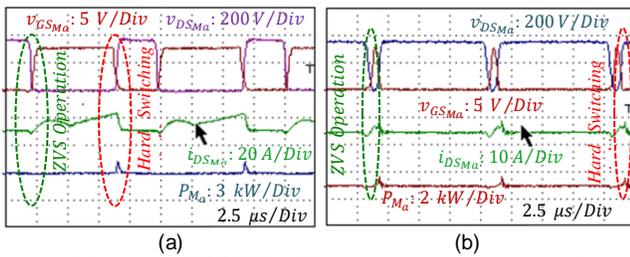
(a) (b)
Fig. 10. $M_a$ power MOSFET different voltage, current, and power losses waveforms under the two worst-case conditions.

Fig. 10 illustrates ZVS operation of the converter even under these worst-case conditions, when the power MOSFETs are turned on. Although, these devices are turned off under the hard switching conditions, but their switching losses are much smaller than the PWM converters ones. Consequently, low EMI noise and high efficiency are achieved. Also, the converter input, output, and $M_a$ drain-source voltages are given in Fig. 11 under these two worst-case conditions.

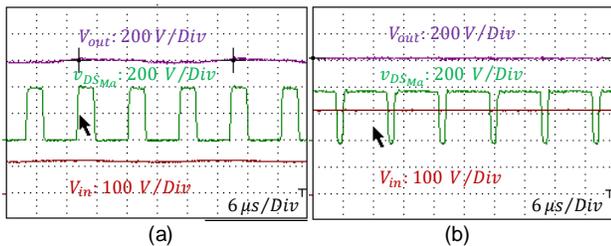
(a) (b)
Fig. 11. converter input, output, and $M_a$ drain-source voltages under the two worst-case conditions.

It must be mentioned that all power diodes are also operate under the soft switching conditions. For instance, different voltages, currents, and power losses waveforms of the two power diodes are given in Figs. 12 and 13 under the abovementioned two worst-case conditions. As can be seen, soft switching operation of these power diodes are realized, even when wide input voltage and load variations are applied. Fig. 14 shows different power MOSFETs and diodes power losses curves versus output power at different input voltages when $V_{out}$=1 kV and 120 V is applied to the converter.

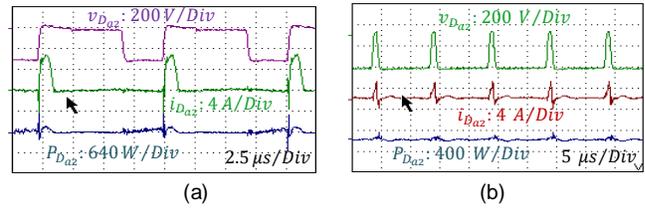
(a) (b)
Fig. 12. $D_{a2}$ power diode different voltage, current, and power losses waveforms under the two worst-case conditions.

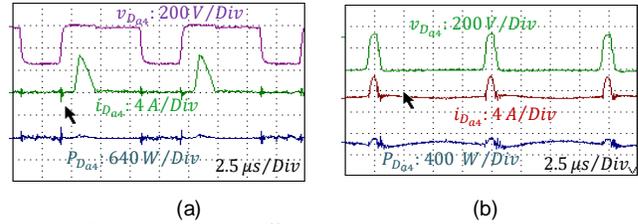
(a) (b)
Fig. 13. $D_{a4}$ power diode different voltage, current, and power losses waveforms under the two worst-case conditions.

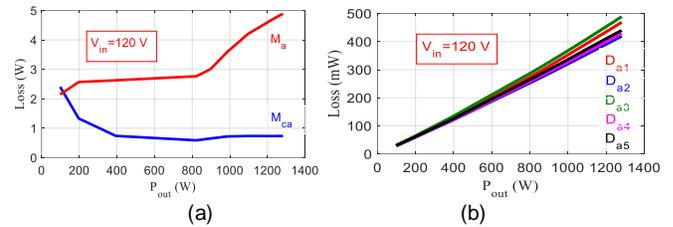
(a) (b)
Fig. 14. Different components losses curves versus output power at different input voltages when $V_{out}$=1 kV (a) MOSFETs and (b) Diodes.

Finally, Fig. 15 shows the converter different efficiency curves versus output power at different input voltages when $V_{out}$=1 kV. Here, the given experimental results have been achieved by changing the MOSFETs duty cycles by tray and error in an open-loop control system to verify the given simulations and analyses. Although, lots of analyses have been given here, but we still have not introduced the converter small-signal model to study its dynamic behavior by using a proper closed-loop control system. This subject and the converter design procedure need more investigations and they are not addressed here to shorten the discussion.

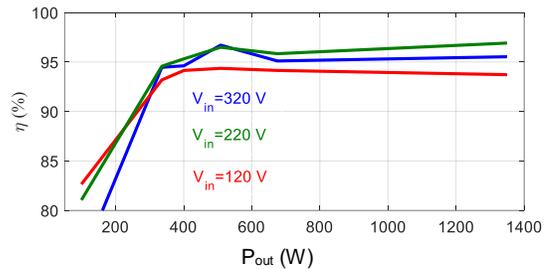
Fig. 15. Converter different efficiency curves versus output power at different input voltages when $V_{out}$=1 kV.

The given converter can realize high voltage gain and it is suitable for high-voltage, high-power, wide-input voltage and wide-load variations applications. Also, low currents and voltages stresses are applied on the converter different components. The prototype converter has been tested under wide input voltage (120-320 V) variation at 100 kHz with maximum efficiency of 96.5%. Although, load variations of 50-1300 W are applied here, due to the laboratory instruments power limitations, but it can be used for higher output power values, as well, and it is a good choice for these applications.

TABLE III COMPARING THE PROPOSED CONVERTER WITH SOME DIFFERENT EXISTING TOPOLOGIES

| | References | $V_{out}$ (V) | $V_{in}$ (V) | $P_{out}$ (W) | Suitable for High Output Voltage | Suitable for High Output Power | Switching Condition | Switching Freq. (kHz) | Switches | Diodes | Main Capacitors Capacitances (µF) | Main Inductors or Transformers Inductances (µH) | Efficiency (%) | Coupled-Inductor or Transformer | In/Out Common Ground |
|---|---|---|---|---|---|---|---|---|---|---|---|---|---|---|---|
| Cockcroft-Walton VM Circuits | [42] | 160k | 600 | 8k | Yes | Yes | ZVS | 300 | 4 | 112 | 112×1n, 14×1.7n | Not Given | 82 | 14 | No |
| | [43] | 1200 | 110 rms | 500 | Yes | Yes | Hard | 60 | 8 | 6 | 6×470 | 1500 | 90.3 | 0 | No |
| | [44] | 450 | 42-54 | 200 | Yes | Yes | Hard | 60 | 4 | 6 | 6×470 | 1500 | 91 | 0 | No |
| | [45] | 250 | 25-30 | 250 | No | No | Hard | 100 | 2 | 5 | 2×(60, 30), 22 | 100, 100 | 95 | 0 | No |
| High-Voltage Gain Interleaved Converters | [46] | 700 | 100 | 1000 | No | Yes | Hard | 10 | 2 | 4 | 2×40, 195 | 1158, 1158 | 97.4 | 0 | Yes |
| | [47] | 800 | 32 | 400 | No | Yes | Hard | 118 | 2 | 7 | 2×10 | 2× (94, 846) | 96.7 | 2 | No |
| | [48] | 675 | 45 | 1000 | Yes | Yes | Hard | 20 | 2 | 8 | 4×60, 3×220 | 2× (320, 320), 2×(700, 700) | 97.8 | 3 | Yes |
| | [49] | 400 | 30-100 | 800 | No | Yes | Hard | 20 | 8 | 0 | 6×270 | 3×350 | 95.8 | 0 | Yes |
| | [50] | 600 | 50 | 1300 | No | Yes | Hard | 20 | 2 | 6 | 4×60, 470 | 3×(320:320), 700:700:700 | 97.7 | 3 | Yes |
| | [51] | 380 | 48 | 3500 | No | Yes | Hard | 50 | 2 | 4 | 2×10, 120 | Not Given | 97.3 | 1 | Yes |
| | [52] | 380 | 48 | 2000 | Yes | Yes | Hard | 50 | 2 | 6 | 4×120 | 2×(80:80:80) | 96.5 | 2 | Yes |
| | [53] | 300 | 20 | 300 | No | No | Hard | 50 | 2 | 6 | 6×15 | 2×200 | 93.6 | 0 | Y/N |
| | [54] | 400 | 42-54 | 1000 | No | Yes | Hard | 25 | 2 | 8 | 6×2.2, 680 | 70µH+ Transformer | 97.5 | 1 | Yes |
| | [55] | 400 | 40-50 | 400 | No | No | ZVS/ ZCS | 40 | 2 | 6 | 6×220 | 2× (150:150) | 96.5 | 2 | No |
| | [56] | 760 | 40-56 | 500 | No | No | ZVS | 100 | 2 | 4 | 4.4, 2×2.2, 2×220 | 88: 460 | 96.5 | 1 | No |
| | [57] | 500 | 35 | 1000 | No | Yes | ZVS | 40 | 4 | 2 | 30, 10, 220, 470 | 2×160, 800: 800 | 97.8 | 1 | Yes |
| | [58] | 270 | 15-30 | 1000 | No | Yes | ZVS | 50 | 4 | 6 | 5×1, 2×47 | 2×(100:100, 2×2) | 97.2 | 2 | Yes |
| | [59] | 380 | 40 | 1000 | No | Yes | ZCS | 100 | 2 | 4 | 4.7, 47, 470 | 2×(95:95) | 94.7 | 2 | Yes |
| | [60] | 380 | 48 | 1500 | No | Yes | ZVS | 50 | 6 | 0 | 2×240, 2×300 | 2×253, 660:2138 | 95.3 | 1 | No |
| | [61] | 380 | 36 | 500 | No | Yes | ZVS | 100 | 2 | 2 | 4.4, 9.4, 470 | 140, 850:4168 | 96.7 | 0 | Yes |
| | [62] | 900 | 59 | 415 | Yes | Yes | ZVS | 42 | 4 | 6 | 4×47, 3×220 | 2× (320: 320: 320) | 96.8 | 2 | Yes |
| Resonant SCC | [63] | 400 | 25-40 | 1000 | No | Yes | ZVS | 100 | 4 | 7 | 7×2, 2×40 | 20, 2×2.7, 3.6, 1.3 | 97.9 | 0 | No |
| | [64] | 190 | 50 | 300 | No | No | ZVS | 35 | 4 | 6 | 3×1.47, 2×4.7 | 3×1.27 | 93.5 | 0 | No |
| | [65] | 60 | 19-21 | 200 | No | Yes | ZVS | >193 | 4 | 4 | 2×0.22, 2×$C_{out}$ | 1.5 | 96.6 | 0 | No |
| | [66] | 21-32 | 100 | 140 | No | Yes | ZVS | 100-200 | 4 | 2 | 2×1, 10, 68 | 2×2.5 | 96.2 | 0 | Yes |
| Proposed Converter | | 1000 | 120-320 | 1300 | Yes | Yes | ZVS | 100 | 4 | 10 | 4×0.33, 3×100 | 2×5, 2×50 | 96.5 | 0 | Yes |

Also, all power devices are operating under the soft switching conditions, as can be concluded from the given simulation and experimental results. Consequently, high switching frequency operation is possible to employ small passive components to achieve high power density. Here, a 100 kHz prototype converter has been implemented due to the laboratory limitations and available components, but high switching frequency operation is possible, if it is necessary.

Finally, in Table III an implemented 3-stage 2-leg configuration of the proposed converter has been compared with some other existing Cockcroft-Walton based VM circuits, high voltage-gain interleaved converters, and some resonant switched-capacitor converters, in details. From this table, some conclusions can be made simply as follows:

a) Although, [42] has been proposed for extra high-voltage, high-power applications, but it uses lots of active and passive components, including massive transformers. So, its reliability, efficiency, and power density are low. Also, other Cockcroft-Walton based VM circuits [43]- [45] are suitable for high-voltage high-power applications. But, they generally operate under the hard switching conditions. So, they have low power densities, too. Also, there is no common ground between their input and output ports.

b) Some high voltage-gain interleaved converters [46]- [54] operate under the hard switching conditions and they are not suitable for low-noise high-power density applications. Although, [55] and [56] operate under the soft switching conditions, but they are not suitable for high-voltage high-power applications and there are no common grounds between their input and output ports, too. Although, [57]- [61] can be used for high-power applications, but they are also not suitable for high-voltage applications. Although, [62] operates under the ZVS conditions and it is a good candidate for high-voltage high-power applications, but to achieve a high voltage-gain value it needs two massive coupled inductors that reduce its power density, in practice.

c) Resonant SCCs [63]- [66] operate under the soft switching conditions and they can provide high-power densities, practically. But, they are not suitable for high-voltage high-power applications. These SCCs, except [66], have no common ground between their input and output ports, too.

d) All power switches and diodes of the proposed converter operate under the soft switching conditions. So, high switching frequency is possible to use small passive components. Also, its symmetrical topology and its common input-output ground feature simplify its input/output filters, as well as its gate driver and control circuits. Some other features such as its transformer-less or coupled-inductor less configuration, multi-leg capability, and its passive and active devices low currents and voltages stresses, make it as a good candidate for low-noise, high-voltage, high-power, and high-power density applications. The main drawback of the given topology is its high number of employed diodes which is not a major issue in practice, because low voltage and current stresses are applied on these components. Consequently, low-cost diodes can be used, practically.

By considering proper stage-count and leg-count values, higher output voltage or output power value can easily achieve by applying reasonable voltage and current stresses on the passive and active components. By increasing the interleaved legs count, the input filter as well as the output capacitors volumes can be reduced effectively to satisfy the desired input current and output voltage ripples values. Also, for the given input-output voltages and power values, different components voltages and currents stresses can be optimized by properly choosing the converter stages- and legs-counts, respectively. This topology can also be used for multi-input applications, which it is not addressed here to shorten the discussion.

## XI. CONCLUSION

A family of current-fed high step up converters by using voltage multiplier, active clamp, and interleaved techniques have been introduced and analyzed mathematically for high-voltage and high-power applications. Their input and output ports common ground feature significantly simplifies the gate-drives and control circuits. Also, low input current and output voltage ripples values, and high voltage-gain characteristics of the proposed converter makes it suitable for lots of dc-dc applications. Simulations and experimental results show that all power devices entirely operate under the soft switching conditions, even when wide load and input voltage variations are applied to the converter. So, high switching frequency operation is possible to achieve high power density. Finally, to validate the analyses and simulations, a 120-320 V to 1 kV, 50-1300 W three-stage two-leg prototype converter has been implemented at 100 kHz with maximum efficiency of 96.5%.

## APPENDIX

### Analysis of the Converter Different Operational States

We are interesting to derive this converter equivalent circuits describing equations during its different operational states in Laplace-domain instead of time-domain, because they can be solved in canonical forms, easily. The following Laplace transforms are needed to analyze each simplified equivalent circuit during the given different time subintervals:

a) a well-known relation,

$$L\left[\frac{d^n f(t)}{dt^n} u(t)\right] = s^n L[f(t)] - s^{n-1} \frac{d^{n-1} f(0)}{dt^{n-1}} - s^{n-2} \frac{d^{n-2} f(0)}{dt^{n-2}} - \cdots - f(0) \quad (A1)$$

b) another well-known time shifted function is also needed as follows:

$$L\left[\frac{d^n f(t-t_i)}{dt^n} u(t-t_i)\right] = e^{-st_i} L\left[\frac{d^n f(t)}{dt^n} u(t)\right] \quad (A2)$$

c) Also, it is necessary to calculate the following Laplace transform during some subintervals:

$$L\left[\frac{d^n f(t)}{dt^n} u(t-t_i)\right]$$

$$= \int_0^\infty \frac{d^n f(t)}{dt^n} e^{-st} dt - \int_0^{t_i} \frac{d^n f(t)}{dt^n} e^{-st} dt$$

$$= L\left[\frac{d^n f(t)}{dt^n} u(t)\right] - \int_0^{t_i} \frac{d^n f(t)}{dt^n} e^{-st} dt \quad (A3)$$

Where, the latest term can be expressed as follows, too:

$$\int_0^{t_i} \frac{d^n f(t)}{dt^n} e^{-st} dt$$

$$= e^{-st} \frac{d^n f(t)}{dt^n}\bigg|_0^{t_i} + s \int_0^{t_i} \frac{d^n f(t)}{dt^n} e^{-st} dt \quad (A4)$$

Therefore,

$$\int_0^{t_i} \frac{d^n f(t)}{dt^n} e^{-st} dt = -\frac{1}{s-1} e^{-st} \frac{d^n f(t)}{dt^n}\bigg|_0^{t_i} \quad (A5)$$

Substituting (A5) into (A3), we can write:

$$L\left[\frac{d^n f(t)}{dt^n} u(t-t_i)\right]$$

$$= L\left[\frac{d^n f(t)}{dt^n} u(t)\right] + \frac{e^{-st}}{s-1} \frac{d^n f(t)}{dt^n}\bigg|_0^{t_i} \quad (A6)$$

Considering (A2) and (A6), and deriving each subinterval equations lead to long expressions which cannot be solved, easily. It must be mentioned that when $f(t)$ is a periodic function with period of $T_s$, then $1/(1 - e^{-sT_s})$ factor must be considered, too. Therefore, to simplify the equations and to avoid boring algebraic calculations, any time-shift during each subinterval, as well as the abovementioned factor, are ignored, first. Consequently, exponential terms in (A2) and (A6) are ignored which simplify the analyses. Then, the simplified equations are solving in the Laplace domain and they are converting into the time-domain by applying the inverse Laplace transforms. Finally, the related time-shifts can be applied correctly for the different time subintervals.

Considering the abovementioned note by ignoring the exponential terms in the equations, and considering the given key waveforms and different operational states, as respectively illustrated in Figs. 3 and 4, the converter is mathematically analyzed, here. During each operational state, two simple closed loops, as shown in different parts of Fig. 4, can be considered to extract the main equations that describe the converter behavior. Actually, there is no any track between the inductors common point and the output port, in some cases. But, a very small parasitic capacitance can be considered between this common point and the output port to generally solve this problem and to extract these two main equations, straightforwardly. Here, only the CCM operation of the converter is analyzed by considering Figs. 3 and 4. It must be mentioned that during each operational state the inductors currents are two main waveforms that must be identified, because the other parameters can be identified for the given inductors currents waveforms, easily. Therefore, only calculating these two key waveforms are given, here.

Considering the converter equivalent circuit during the first operational state, as shown in Fig. 4 (a), and applying the KVL on loops (1) and (2) in time-domain we can write:

$$L_a \frac{di_{L_{a1}}(t)}{dt} + L_{ra} \frac{di_{L_{ra1}}(t)}{dt} = v_{in}(t) \quad (A7)$$

$$L_{ra} \frac{di_{L_{ra1}}(t)}{dt} = v_{C_{out1}}(t) \quad (A8)$$

Here, $i_{L_{rai}}(t)$ and $i_{L_{ai}}(t)$ terms have been used to identify the $L_a$ and $L_{ra}$ inductors currents during $i^{th}$ operational states, generally. Where, $i \in \{1, 2, 3, \ldots, 8\}$ shows the $i^{th}$ operational state. (A7) and (A8) can be rewritten in Laplace-domain in matrix form by considering $i_{L_a}(t_1)$ and $i_{L_{ra}}(t_1)$ initial values:

$$\begin{bmatrix} L_a s & L_{ra} s \\ 0 & L_{ra} s \end{bmatrix} \begin{bmatrix} I_{L_{a1}}(s) \\ I_{L_{ra1}}(s) \end{bmatrix} = \begin{bmatrix} V_{in}(s) + L_a i_{L_a}(t_1) + L_{ra} i_{L_{ra}}(t_1) \\ V_{C_{out1}}(s) + L_{ra} i_{L_{ra}}(t_1) \end{bmatrix} \quad (A9)$$

During this operational state, $i_{L_{ra}}$ is continuously increased. Finally, $i_{L_{ra}}(t_2) = i_{L_a}(t_2)$ and $D_{a1}$ is turned off under the ZCS condition and this operational state is finished. According to Fig. 3, $i_{L_{ra}}$ is more increased and $D_{a4}$ is consequently turned on under the ZCS condition without any switching losses.

Also, during the second operational state by considering the converter equivalent circuit, as shown in Fig. 4 (b), and applying the KVL on loops (1) and (2), we can write:

$$L_a \frac{di_{L_{a2}}(t)}{dt} + L_{ra} \frac{di_{L_{ra2}}(t)}{dt} = v_{in}(t) \quad (A10)$$

$$L_{ra} \frac{di_{L_{ra2}}(t)}{dt} + v_{C_{a1}}(t) + v_{C_{a2}}(t) = v_{C_{out1}}(t) + v_{C_{out2}}(t) \quad (A11)$$

During this operational state, we can write:

$$i_{C_{a1}}(t) = i_{C_{a2}}(t) = i_{L_{ra2}}(t) - i_{L_{a2}}(t) \quad (A12)$$

Substituting (A12) into (A11), considering state-space initial values, and doing some straightforward algebraic calculations, (A10) and (A11) can be rewritten in Laplace domain, as follows:

$$\begin{bmatrix} L_a s & L_{ra} s \\ -\left(\frac{1}{C_{a1}} + \frac{1}{C_{a2}}\right)\frac{1}{s} & L_{ra} s + \left(\frac{1}{C_{a1}} + \frac{1}{C_{a2}}\right)\frac{1}{s} \end{bmatrix} \begin{bmatrix} I_{L_{a2}}(s) \\ I_{L_{ra2}}(s) \end{bmatrix} =$$

$$\begin{bmatrix} V_{in}(s) + L_a i_{L_a}(t_2) + L_{ra} i_{L_{ra}}(t_2) \\ \left(-v_{C_{a1}}(t_2) - v_{C_{a2}}(t_2)\right)\frac{1}{s} + V_{C_{out1}}(s) + V_{C_{out2}}(s) \end{bmatrix} \quad (A13)$$

During this time subinterval, $v_{C_{a2}}$ is continuously increased, but $V_{C_{out2}}$ is approximately constant, because $C_{a2}$ has been chosen much smaller than $C_{out2}$. Consequently, $v_{C_{a2}}(t_3) = V_{C_{out2}}$ and $D_{a2}$ is turned on under the ZVS condition. Then, both $D_{a2}$ and $D_{a4}$ conduct current during a short time interval and $D_{a4}$ is finally reversed biased and this operational state is terminated. Generally, we can assume $V_{C_{out1-3}} \approx V_{C_{out}}/m$.

In the same manners and by considering Fig. 4, the other operational states describing equations can be deriving out easily by applying KVL on the two loops, identified on the converter equivalent simplified circuits. For instance, state V starts when the controller turns $M_a$ off. The switching losses are small due to the presence of $C_{ra}$ in parallel with this power MOSFET. Neither the power MOSFETs nor the diodes conduct current during this time interval, as shown in Figs. 3, and 4 (e). Actually, $D_{a5}$ is turned on after a short time later during this subinterval, but to simplify the analysis this diode has been considered off. Applying KVL on the both identified loops on Fig. 4 (e), and doing some straightforward algebraic calculations, as done before, the converter equivalent circuit describing equations can be given in matrix form in Laplace domain as follows:

$$\begin{bmatrix} L_a s & L_{ra} s + \frac{1}{C_{ra}}\frac{1}{s} \\ -\frac{1}{C_p}\frac{1}{s} & L_{ra} s + \left(\frac{1}{C_p} + \frac{1}{C_{ra}}\right)\frac{1}{s} \end{bmatrix} \begin{bmatrix} I_{L_{a5}}(s) \\ I_{L_{ra5}}(s) \end{bmatrix} =$$

$$\begin{bmatrix} \left(-v_{C_{ra}}(t_5)\right)\frac{1}{s} + V_{in}(s) + L_a i_{L_a}(t_5) + L_{ra} i_{L_{ra}}(t_5) \\ \left(-v_{C_{ra}}(t_5) - v_{C_p}(t_5)\right)\frac{1}{s} + V_{out}(s) + L_{ra} i_{L_{ra}}(t_5) \end{bmatrix} \quad (A14)$$

Deriving the other operational states equations is straightforward and it is ignored here to shorten the discussion.

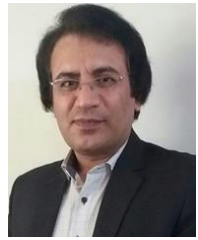

**Reza Beiranvand** (SM'21) received the MSc and PhD degrees in electrical engineering from Sharif University of Technology, Tehran, Iran, in 1999 and 2010, respectively. From 2010 to 2012, he was a Postdoctoral research fellow with the Faculty of Electrical Engineering, Sharif University of Technology. From 1999 to 2007, he was an engineer at the R&D centers of PARS-Electric and RADIO SHAHAB MFGs, Tehran, Iran, where he was engaged in designing the LCD, and LED TVs based on the ST, LT, NXP, and Fairchild devices.

He is currently an Associate Professor at the Faculty of Electrical and Computer Engineering, Tarbiat Modares University, Tehran, Iran.

Dr. Beiranvand was the IEEE Consultant (2017-2019) and Head of the Power Group (2018-2020) in Tarbiat Modares University. He is among the "Top 2 Percent Scientists of the World", based on what is called Stanford University Released List in Elsevier, 2020 and 2021. His research interests include power electronics converters, soft switching techniques, SCCs, SMPS, Capacitive-Coupling Power Transfer (CPT) and Inductive Power Transfer (IPT) techniques, and PV-based renewable energy systems.

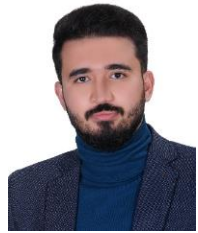

**Soheil Hasani Sangani** was born in Iran in 1995. He received the BSc. degree in power engineering from Ferdowsi University, Mashhad, Iran, in 2017, and the MSc. degree in power electronics engineering from Tarbiat Modares University, Tehran, Iran, in 2020.

His current research interests include power electronics, high step-up dc-dc converters, soft switching techniques, and PV-based renewable energy systems.